\documentclass[prb,aps,twocolumn,showpacs,floats]{revtex4}
\usepackage{epsfig,bm,epsf,graphics}
\begin{document}
\draft
\title{Critical behavior of magnetic thin films as a function of thickness}
\author{X. T. Pham Phu$^{a}$, V. Thanh Ngo$^{b,c}$ and H. T.
Diep$^{a}$\footnote{ Corresponding author, E-mail:diep@u-cergy.fr }}
\address{$^{a}$ Laboratoire de Physique Th\'eorique et Mod\'elisation,
CNRS-Universit\'e de Cergy-Pontoise, UMR 8089\\
2, Avenue Adolphe Chauvin, 95302 Cergy-Pontoise Cedex, France\\
$^b$ Institute of Physics, P.O. Box 429,   Bo Ho, Hanoi 10000,
Vietnam\\
$^c$ Asia Pacific Center for Theoretical Physics, Hogil Kim
Memorial Building 5th floor, POSTECH, Hyoja-dong, Namgu, Pohang
790-784, Korea}

\begin{abstract}
We study  the critical behavior of magnetic thin films as a
function of the film thickness.  We use the ferromagnetic Ising
model with the high-resolution multiple histogram Monte Carlo (MC)
simulation. We show that though the 2D behavior remains dominant
at small thicknesses, there is a systematic continuous deviation
of the critical exponents from their 2D values. We observe that in
the same range of varying thickness the deviation of the exponent
$\nu$ is rather small, while exponent $\beta$ suffers a larger
deviation.  We explain these deviations using the concept of
"effective" exponents suggested by Capehart and Fisher in a
finite-size analysis.  The shift of the critical temperature with
the film thickness obtained here by MC simulation is in an
excellent agreement with their prediction.
\end{abstract}

 \maketitle

\section{Introduction}

During the last 30 years.  physics of surfaces and objects of
nanometric size have attracted an immense interest.  This is due
to important applications in
industry.\cite{zangwill,bland-heinrich}
 An example is the so-called giant
magneto-resistance (GMR) used in data storage devices, magnetic
sensors, ... \cite{Baibich,Grunberg,Fert,review}  In parallel to
these experimental developments, much theoretical
effort\cite{Binder-surf,Diehl} has also been devoted to the search
of physical mechanisms lying behind new properties found in
nanoscale objects such as ultrathin films, ultrafine particles,
quantum dots, spintronic devices etc. This effort aimed not only
at providing explanations for experimental observations but also
at predicting new effects for future experiments.

The physics of two-dimensional (2D) systems is very exciting. Some
of those 2D systems can be exactly solved: one famous example is
the Ising model on the square lattice which has been solved by
Onsager.\cite{Onsager}  This model shows a phase transition at a
finite temperature $T_c$ given by $\sinh^2(2J / k_BT_c)=1$ where
$J$ is the nearest-neighbor (NN) interaction. Another interesting
result is the absence of long-range ordering at finite
temperatures for the continuous spin models (XY and Heisenberg
models) in 2D.\cite{Mermin} In general, three-dimensional (3D)
systems for any spin models cannot be unfortunately solved.
However, several methods in the theory of phase transitions and
critical phenomena can be used to calculate the critical behaviors
of these systems.\cite{Zinn}

This paper deals with systems between 2D and 3D. Many theoretical
studies have been devoted to thermodynamic properties of thin films,
magnetic multilayers,...
\cite{Binder-surf,Diehl,ngo2004trilayer,Diep1989sl,diep91-af-films}
In spite of this, several points are still not yet understood.   It
is known a long time ago that the presence of a surface in magnetic
materials can give rise to surface spin-waves which are localized in
the vicinity of the surface.\cite{diep79} These localized modes may
be acoustic with a low-lying energy or optical with a high energy,
in the spin-wave spectrum.  Low-lying energy modes contribute to
reduce in general surface magnetization at finite temperatures. One
of the consequences is the surface disordering which may occur at a
temperature lower than that for interior magnetization.\cite{diep81}
The existence of low-lying surface modes depends on the lattice
structure, the surface orientation, the surface parameters, surface
conditions (impurities, roughness, ...) etc. There are two
interesting cases: in the first case a surface transition occurs at
a temperature distinct from that of the interior spins and in the
second case the surface transition coincides with the interior one,
i. e. existence of a single transition. Theory of critical phenomena
at surfaces\cite{Binder-surf,Diehl} and Monte Carlo (MC)
simulations\cite{Landau1,Landau2} of critical behavior of the
surface-layer magnetization at the extraordinary transition in the
three-dimensional Ising model have been carried out.  These works
suggested several scenarios in which the nature of the surface
transition and the transition in thin films depends on many factors
in particular on the symmetry of the Hamiltonian and on surface
parameters.

We confine ourselves here in the case of a simple cubic film with
Ising model. For our purpose, we suppose all interactions are the
same everywhere even at the surface.
 This case is the simplest case where there is no surface-localized
spin-wave
 modes and there is only a single phase transition at a temperature for the
 whole system (no separate surface phase transition).\cite{diep79,diep81}
  Other complicated cases will
 be left for future investigations.  However, some preliminary discussions
 on this point for complicated surfaces
  have been reported
 in some of our previous papers.\cite{ngo2007,ngo2007fcc}
In the case of a simple cubic film with Ising model,  Capehart and
Fisher have studied the critical behavior of the susceptibility
using a finite-size scaling analysis.\cite{Fisher} They showed that
there is a crossover from 2D to 3D behavior as the film thickness
increases.  The so-called "effective" exponent $\gamma$ has been
shown to vary according to a scaling function depending both on the
film thickness and the distance to the transition temperature. As
will be seen below the scaling suggested by Capehart and Fisher is
in agreement with what we find here using extensive MC simulation.

The aim of this paper is  to investigate the effect of the thickness
on the critical exponents of the film.
 To carry out these
purposes, we shall use MC simulations with highly accurate
multiple histogram technique.\cite{Ferrenberg1,Ferrenberg2,Bunker}

The paper is organized as follows. Section II is devoted to a
description of the model and method.   Results are shown and
discussed in section III. Concluding remarks are given in section
IV.

\section{Model and Method}
\subsection{Model} Let us consider the Ising spin model on a film
made from a ferromagnetic simple cubic lattice. The size of the
film is $L\times L\times N_z$.  We apply the periodic boundary
conditions (PBC) in the  $xy$ planes to simulate an infinite $xy$
dimension. The $z$ direction is limited by the film thickness
$N_z$.
  If $N_z=1$ then one has a 2D square lattice.

The Hamiltonian is given by
\begin{equation}
\mathcal H=-\sum_{\left<i,j\right>}J_{i,j}\sigma_i\cdot\sigma_j
\label{eqn:hamil1}
\end{equation}
where $\sigma_i$ is the Ising spin of magnitude 1 occupying the
lattice site $i$, $\sum_{\left<i,j\right>}$ indicates the sum over
the NN spin pairs  $\sigma_i$ and $\sigma_j$.

In the following, the interaction between two NN surface spins is
denoted by $J_s$, while all other interactions are supposed to be
ferromagnetic and all equal to $J=1$ for simplicity. Let us note
in passing that in the semi-infinite crystal  the surface phase
transition occurs at the bulk transition temperature when $J_s
\simeq 1.52 J$. This point is called "extraordinary phase
transition" which is characterized by some particular critical
exponents.\cite{Landau1,Landau2}   In the case of thin films, i.
e. $N_z$ is finite, it has been theoretically shown that when
$J_s=1$ the bulk behavior is observed when the thickness becomes
larger than a few dozens of atomic layers:\cite{diep79} surface
effects are insignificant on thermodynamic properties such as the
value of the critical temperature, the mean value of magnetization
at a given $T$, ... When $J_s$ is smaller than $J$, surface
magnetization is destroyed at a temperature lower than that for
bulk spins.\cite{diep81} The criticality of a film with uniform
interaction, i.e. $J_s=J$, has been studied by Capehart and Fisher
as a function of the film thickness using a scaling
analysis\cite{Fisher} and by MC simulations.\cite{Schilbe,Caselle}
The results by Capehart and Fisher indicated that as long as the
film thickness is finite the phase transition is strictly  that of
the 2D Ising universality class.  However, they showed that at a
temperature away from the transition temperature $T_c(N_z)$, the
system can behave as a 3D one when the spin-spin correlation
length $\xi(T)$ is much smaller than the film thickness, i. e.
$\xi(T)/N_z\ll 1$. As $T$ gets very close to $T_c(N_z)$,
$\xi(T)/N_z \rightarrow 1$, the system undergoes a crossover to 2D
criticality.  We will return to this work for comparison with our
results shown below.

\subsection{Multiple histogram technique}\label{mht}

The multiple histogram technique is known to reproduce with very
high accuracy the critical exponents of second order phase
transitions.\cite{Ferrenberg1,Ferrenberg2,Bunker}

The overall probability distribution\cite{Ferrenberg2} at
temperature $T$ obtained from $n$ independent simulations, each
with $N_j$ configurations, is given by
\begin{equation}
P(E,T)=\frac{\sum_{i=1}^n H_i(E)\exp[E/k_BT]}{\sum_{j=1}^n
N_j\exp[E/k_BT_j-f_j]}, \label{eq:mhp}
\end{equation}
where
\begin{equation}
\exp[f_i]=\sum_{E}P(E,T_i). \label{eq:mhfn}
\end{equation}

The thermal average of a physical quantity $A$ is then calculated
by
\begin{equation}
\langle A(T)\rangle=\sum_E A\,P(E,T)/z(T),
\end{equation}
in which
\begin{equation}
z(T)=\sum_E P(E,T).
\end{equation}

Thermal averages of physical quantities are thus calculated as
continuous functions of $T$, now the results should be valid over
a much wider range of temperature than for any single histogram.

In MC simulations, one calculates the averaged order parameter
$\langle M\rangle$ ($M$: magnetization of the system), averaged
total energy $\langle E\rangle$, specific heat $C_v$,
susceptibility $\chi$, first order cumulant of the energy $C_U$,
and $n^{th}$ order cumulant of the order parameter $V_n$ for $n=1$
and 2. These quantities are defined as

\begin{eqnarray}
\langle E\rangle&=&\langle\cal{H}\rangle,\\
C_v&=&\frac{1}{k_BT^2}\left(\langle E^2\rangle-\langle E\rangle^2\right),\\
\chi&=&\frac{1}{k_BT}\left(\langle M^2\rangle-\langle M\rangle^2\right),\\
C_U&=&1-\frac{\langle E^4\rangle}{3\langle E^2\rangle^2},\\
V_n&=&\frac{\partial\ln{M^n}}{\partial(1/k_BT)} =\langle
E\rangle-\frac{\langle M^nE\rangle}{\langle M^n\rangle}.
\end{eqnarray}

Let us discuss the case where all dimensions can go to infinity.
For example, consider a system of size $L^d$ where $d$ is the
space dimension.  For a finite $L$, the pseudo "transition"
temperatures can be identified by the maxima of $C_v$ and $\chi$,
.... These maxima do not in general take place at the same
temperature. Only at infinite $L$ that the pseudo "transition"
temperatures of these respective quantities coincide at the real
transition temperature $T_c(\infty)$. So when we work at the
maxima of $V_n$, $C_v$ and $\chi$, we are in fact working at
temperatures away from $T_c(\infty)$.   Let us define the reduced
temperature which measures the "distance" from $T_c(\infty)$ by

\begin{equation}\label{rt}
t=\frac{T-T_c(\infty)}{T_c(\infty)}
\end{equation}
This distance tends to zero when all dimensions go to infinity.
For large values of $L$, the following scaling relations are
expected (see details in Ref. \onlinecite{Bunker}):

\begin{equation}\label{V1}
V_1^{\max}\propto L^{1/\nu}, \hspace{1cm} V_2^{\max}\propto
L^{1/\nu},
\end{equation}
\begin{equation}
C_v^{\max}=C_0+C_1L^{\alpha/\nu}\label{Cv}
\end{equation}
and
\begin{equation}\label{chis}
\chi^{\max}\propto L^{\gamma/\nu}
\end{equation}
at their respective 'transition' temperatures $T_c(L)$, and

\begin{equation}
C_U=C_U[T_c(\infty)]+AL^{-\alpha/\nu},
\end{equation}
\begin{equation}
M_{T_c(\infty)}\propto L^{-\beta/\nu}\label{MB}
\end{equation}
and
\begin{equation}
T_c(L)=T_c(\infty)+C_AL^{-1/\nu}\label{TC},
\end{equation}
where $A$, $C_0$, $C_1$ and $C_A$ are constants. We estimate $\nu$
independently from $V_1^{\max}$ and $V_2^{\max}$. With this value
we calculate $\gamma$ from $\chi^{\max}$ and $\alpha$ from
$C_v^{\max}$.  Note that we can estimate $T_c(\infty)$ using the
last expression. Then, using $T_c(\infty)$, we can calculate
$\beta$ from $M_{T_c(\infty)}$. The Rushbrooke scaling law $\alpha
+2\beta +\gamma=2$ is then in principle verified.

Let us emphasize that the expressions Eqs. (\ref{V1})-(\ref{TC})
are valid for large $L$.  To be sure that $L$ are large enough,
one has to allow for corrections to scaling of the form, for
example,

\begin{eqnarray}
\chi^{\max}&=& B_1L^{\gamma/\nu}(1+B_2L^{-\omega})\label{chic}\\
V_n^{\max}&=& D_1L^{1/\nu}(1+D_2L^{-\omega})\label{V1c}
\end{eqnarray}
where $B_1$, $B_2$, $D_1$  and $D_2$ are constants and $\omega$ is a
correction exponent.\cite{Ferrenberg3} Similar forms exist also for
the other exponents.  Usually, these corrections are extremely small
if $L$ is large enough as is the case with today's large-memory
computers. So, in general they do not therefore alter the results
using Eqs. (\ref{V1})-(\ref{TC}).

\subsection{The case of films with finite thickness}\label{film}

In the case of a thin film of size $L\times L\times N_z$, Capehart
and Fisher\cite{Fisher} have showed that as long as the film
thickness $N_z$ is not allowed to go to infinity, there is a 2D-3D
crossover if one does not work at the real transition temperature
$T_c(L=\infty,N_z)$.  Following Capehart and Fisher, let us define

\begin{eqnarray}
\dot{t}&=&\frac{T-T_c(L=\infty, N_z)}{T_c(3D)}\label{rt1}\\
x&=&N_z^{1/\nu_{3D}}\dot{t}
\end{eqnarray}
where $\nu_{3D}$ is the 3D $\nu$ exponent and $T_c(3D)$ the 3D
critical temperature. When $x$ is larger than a value $x_0$, i. e.
at a temperature away from $T_c(L=\infty, N_z)$, the system behaves
as a 3D one. While when $x < x_0$, it should behave as a 2D one.
This crossover was argued from a comparison of the correlation
length in the $z$ direction to the film thickness. As a consequence,
if we work exactly at $T_c(L=\infty, N_z)$ we should observe the 2D
critical exponents for finite $N_z$. Otherwise, we should observe
the so-called "effective critical exponents" whose values are found
between those of 2D and 3D cases.  This point is fundamentally very
important.  There have been some attempts to verify it by MC
simulations,\cite{Schilbe} but these results were not convincing due
to their poor MC quality.  In the following we show with
high-precision MC technique that the prediction of Capehart and
Fisher is really verified.

\section{Results}

The $xy$ linear sizes  $L=20, 24, 30, ...,80$ have been used in our
simulations. For $N_z=3$, sizes up to 160 have been used to evaluate
corrections to scaling.

In practice, we use first the standard MC simulations to localize
for each size the transition temperatures $T^E_0(L)$ for specific
heat and $T^m_0(L)$ for susceptibility. The equilibrating time is
from 200000 to 400000 MC steps/spin and the averaging time is from
500000 to 1000000 MC steps/spin. Next, we make histograms at $8$
different temperatures $T_j(L)$ around the transition temperatures
$T^{E,m}_0(L)$ with 2 millions MC steps/spin, after discarding 1
millions MC steps/spin for equilibrating. Finally, we make again
histograms at $8$ different temperatures around the new transition
temperatures $T^{E,m}_0(L)$ with $2\times 10^6$ and $4\times 10^6$
MC steps/spin for equilibrating and averaging time, respectively.
Such an iteration procedure gives extremely good results for
systems studied so far.  Errors shown in the following have been
estimated using statistical errors, which are very small thanks to
our multiple histogram procedure, and fitting errors given by
fitting software.

We note that only $\nu$ is directly calculated from MC data.
Exponent $\gamma$ obtained from $\chi^{\max}$ and $\nu$ suffers
little errors (systematic errors and errors from $\nu$). Other
exponents are obtained by MC data and  several-step fitting. For
example, to obtain $\alpha$ we have to fit $C_v^{\max}$ of Eq.
\ref{Cv} by choosing $C_0$, $C_1$ and by using the value of $\nu$.
So, in practice, in most cases, one calculates $\alpha$ or $\beta$
from MC data and uses the Rushbrooke scaling law to calculate the
remaining exponent.

Now, similar to the discussion given in subsection \ref{mht}, if we
work at a distance away from $T_c(L=\infty, N_z)$ we should observe
"effective critical exponents".  This is the case because in the
finite size analysis using the multiple histogram technique, we
measure the maxima of $V_n$, $C_V$ and $\chi$ which occur at
different temperatures for a finite $L$. These temperatures, though
close to, are not $T_c(L=\infty, N_z)$. To give a precision on this
point, we show the values of these maxima and the corresponding
temperatures for $N_z=7$ in Table \ref{Nz7}.  For the value of
$T_c(L=\infty, N_z=7)$, see Table \ref{tab:criexp}.

\begin{widetext}
\begin{table*}
  \centering
  \caption{Maxima and temperatures at the maxima of $V_n$ ($n=1,2$), $C_v$
  and $\chi$ for various $L$ with $N_z=7$.}\label{Nz7}
  \begin{tabular}{| r | c | c | c | c | c | c | c |c |}
    \hline
    L & $C_v^{max}$ & $\chi^{max}$ & $V_1^{max}$  & $V_2^{max}$  & $Tc (C_v^{max})$ &
$T_c(\chi^{max})$& $T_c (V_1^{max})$&$T_c (V_2^{max}) $\\
\hline  30 &   2.21115658 &  25.29532589 & 164.05948154&
275.71581036&
4.19027500 &   4.22755000 &   4.24277500&    4.24900000  \\
40   & 2.36517434 &  41.20958927  &219.30094769&  368.72462473&
4.19305000 &   4.21895000  &  4.23025000  &  4.23500000 \\
50  &  2.50496719  & 60.82008190  &275.66203381&  463.17327477&
4.19275000   & 4.21340000 &   4.22210000  & 4.22600000 \\
60&2.59177903 &  82.96529587 & 329.65536262  &554.47606570&
4.19270000 &4.20940000   & 4.21710000   & 4.22045000\\
 70&2.70129995 &109.00528127 & 387.47245040&  651.24905512&
 4.19250000&
4.20640000 &4.21260000 &   4.21530000\\
 80 &   2.76931676&
138.78113065& 443.00488386 & 743.61068938&
4.19220000  &  4.20410000  &  4.20965000 &   4.21205000 \\
    \hline
  \end{tabular}
\end{table*}
\end{widetext}

Given this fact, we emphasize that calculations using Eqs.
(\ref{V1})-(\ref{TC}) will give effective critical exponents except
of course for the case $N_z=1$ where the results correspond to real
critical exponents.

We show now the results obtained by MC simulations with the
Hamiltonian (\ref{eqn:hamil1}).  We have tested that all exponents
do not change in the finite size scaling with $L\geq 30$. So most
of results are shown for $L\geq 30$ except for $\nu$ where the
lowest sizes $L=20,24$ can be used without modifying its value.

\begin{figure} [h!]
\centerline{\epsfig{file=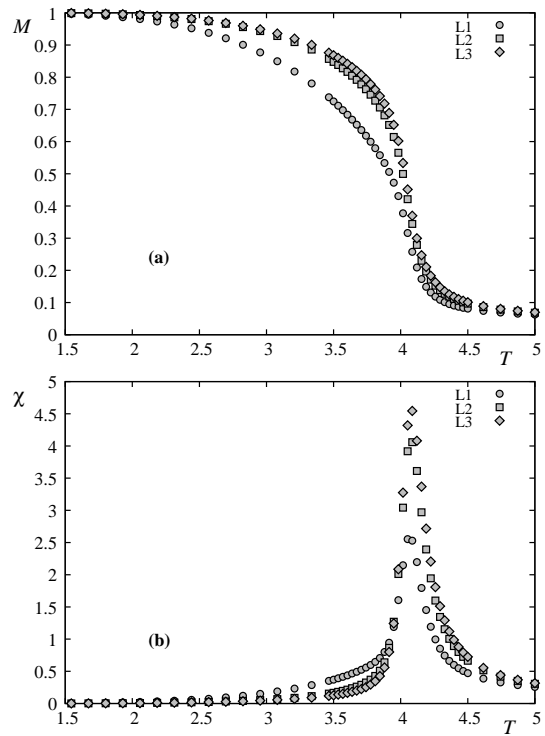,width=2.8in}} \caption{ Layer
magnetizations (a) and layer susceptibilities (b) versus $T$ with
$N_z=5$ and $L=24$.} \label{fig:N24Z5M}
\end{figure}

Let us show in Fig. \ref{fig:N24Z5M}  the layer magnetizations and
their corresponding susceptibilities of the first three layers, in
the case where $J_s=1$.  It is interesting to note that the surface
layer is smaller that the interior layers, as it has been shown
theoretically by the Green's function method a long time
ago.\cite{diep79,diep81}  The surface spins have smaller local field
due to the lack of neighbors, so thermal fluctuations will reduce
more easily the surface magnetization with respect to the interior
ones. The susceptibilities have their peaks at the same temperature,
indicating a single transition.

\begin{figure}
\centerline{\epsfig{file=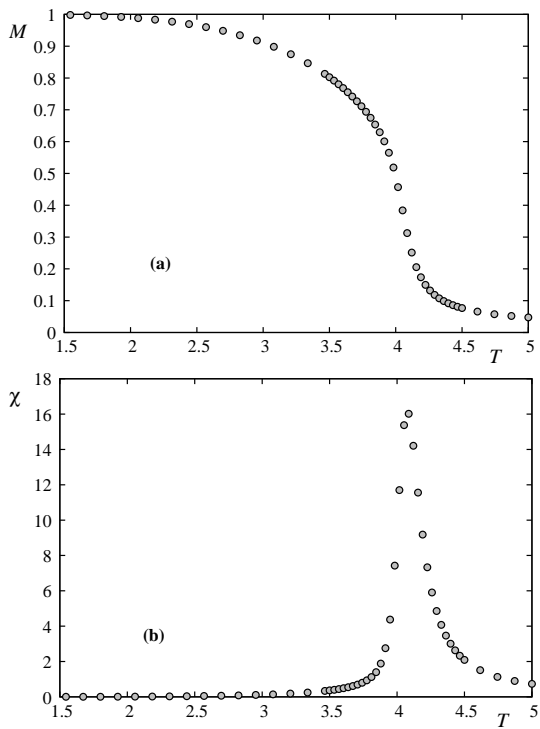,width=2.8in}} \caption{Total
magnetization (a) and total susceptibility (b) versus $T$ with
$N_z=5$ and $L=24$.} \label{fig:N24Z5MT}
\end{figure}

Figure \ref{fig:N24Z5MT} shows total magnetization of the film and
the total susceptibility. This indicates clearly that there is
only one peak as said above.

\subsection{Finite size scaling}

\begin{figure}
\centerline{\epsfig{file=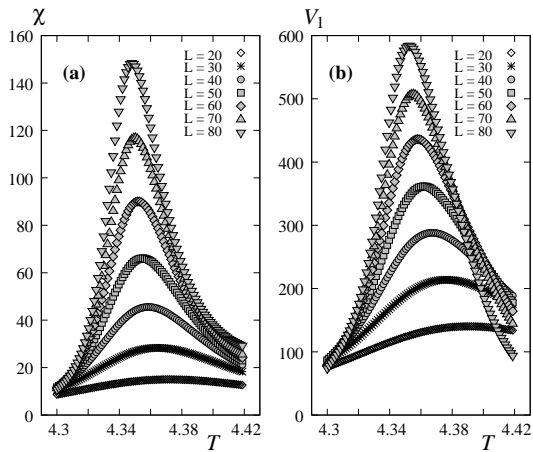,width=2.8in}}
\caption{(a) Susceptibility and (b) $V_1$, as functions of $T$ for
several $L$ with $N_z=11$, obtained by multiple histogram
technique.} \label{fig:ISSVZ11}
\end{figure}

Let us show some results obtained from multiple histograms described
above. Figure \ref{fig:ISSVZ11} shows the susceptibility and the
first derivative $V_1$ versus $T$ around their maxima for several
sizes.

\begin{figure}
\centerline{\epsfig{file=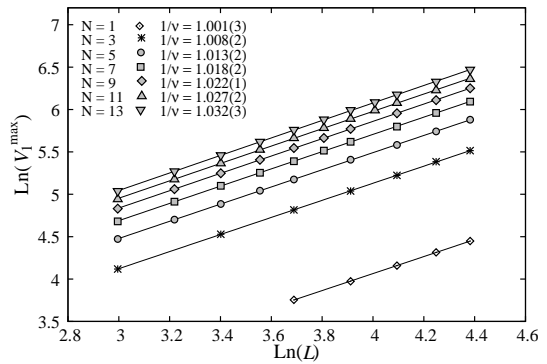,width=2.8in}} \caption{ Maximum
of the first derivative of $\ln M$  versus $ L$ in the $\ln-\ln$
scale.  The  slopes are indicated on the figure. } \label{fig:NUL}
\end{figure}

\begin{figure}
\centerline{\epsfig{file=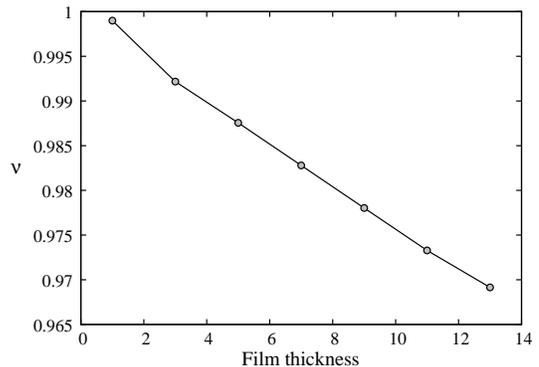,width=2.8in}} \caption{Effective
exponent $\nu$ versus $N_z$.} \label{fig:NUZ}
\end{figure}

We show in Fig. \ref{fig:NUL} the maximum  of the first derivative
of $\ln M$ with respect to $\beta=(k_BT)^{-1}$  versus $ L$ in the
$\ln-\ln$ scale for several film thicknesses up to $N_z=13$. If we
use Eq. (\ref{V1}) to fit these lines, i. e. without correction to
scaling, we obtain $1/\nu$ from the slopes of the remarkably
straight lines.   These values are indicated on the figure.  In
order to see the deviation from the 2D exponent, we plot in Fig.
\ref{fig:NUZ} $\nu$ as a function of thickness $N_z$. We observe
here a small but systematic deviation of $\nu$ from its 2D value
($\nu_{2D}=1)$ with increasing thickness.  To show the precision
of our method, we give here the results of $N_z=1$. For $N_z =1$,
we have $1/\nu =1.0010 \pm 0.0028$ which yields $\nu = 0.9990\pm
0.0031$ and $\gamma/\nu = 1.7537 \pm 0.0034$ (see Figs.
\ref{fig:GAML} and \ref{fig:GAMZ} below) yielding $\gamma =
1.7520\pm 0.0062$. These results are in excellent agreement with
the exact results $\nu_{2D}=1$ and $\gamma_{2D}=1.75$.  The very
high precision of our method is thus verified in the rather modest
range of the system sizes $L=20-80$ used in the present work. Note
that the result of Ref.\onlinecite{Schilbe} gave $\nu=0.96\pm0.05$
for $N_z=1$ which is very far from the exact value.

The deviation of $\nu$ from the 2D value when $N_z$ increases is
due, as discussed earlier, to the crossover to 3D ($\dot {t}$ is
not zero).  Other exponents will suffer the same deviations as
seen below.

We show in Fig. \ref{fig:GAML} the maximum of the susceptibility
versus $L$ in the $\ln-\ln$ scale for  film thicknesses up to
$N_z=13$. We have used only results of $L\geq 30$. Including
$L=20$ and 24 will result, unlike the case of $\nu$,  in a
decrease of $\gamma$ of about one percent for $N_z\geq 7$.  From
the slopes of these straight lines, we obtain the values of
effective $\gamma/\nu$. Using the values of $\nu$ obtained above,
we deduce the values of $\gamma$ which are plotted in Fig.
\ref{fig:GAMZ} as a function of thickness $N_z$. Unlike the case
of $\nu$, we observe here a stronger deviation of $\gamma$ from
its 2D value (1.75) with increasing thickness.  This finding is
somewhat interesting: the magnitude of the deviation from the 2D
value may be different from one critical exponent to another:
$\simeq 3\%$ for $\nu$ and $\simeq 8\%$ for $\gamma$ when $N_z$
goes from 1 to 13. We will see below that $\beta$ varies even more
strongly.

\begin{figure}
\centerline{\epsfig{file=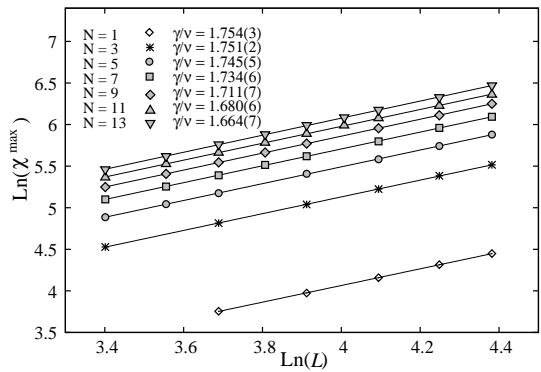,width=2.8in}} \caption{Maximum
of susceptibility versus $L$ in the $\ln-\ln$ scale. The  slopes are
indicated on the figure.} \label{fig:GAML}
\end{figure}

\begin{figure}
\centerline{\epsfig{file=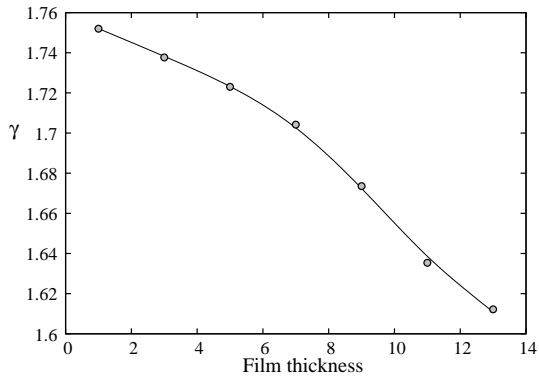,width=2.8in}} \caption{Effective
exponent $\gamma$ versus $N_z$.} \label{fig:GAMZ}
\end{figure}

\begin{figure}
\centerline{\epsfig{file=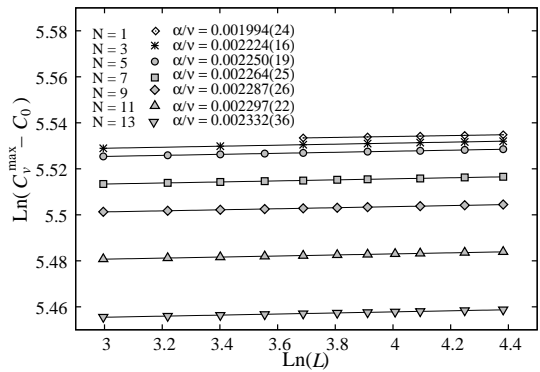,width=2.8in}} \caption{$\ln
 (C_v^{\max}-C_0)$ versus $\ln L$ for
$N_z=1,3,5,...,13$.  The slope gives $\alpha/\nu$  (see Eq.
\ref{Cv}) indicated on the figure.  } \label{fig:ALPHAL}
\end{figure}

We show now in Fig. \ref{fig:ALPHAL} the maximum of $C_v^{\max}$
versus $L$ for $N_z=1,3,5,...,13$.  Note that for each $N_z$ we had
to look for $C_0$, $C_1$ and $\alpha/\nu$ which give the best fit
with data of $C_v^{\max}$. Due to the fact that there are several
parameters which can induce a wrong combination of them, we impose
that $\alpha$ should satisfy the condition $0\leq \alpha \leq 0.11$
where the lower limit of $\alpha$ corresponds to the value of 2D
case and the upper limit to the 3D case.  In doing so, we get very
good results shown in Fig. \ref{fig:ALPHAL}.  From these ratios of
$\alpha/\nu$ we deduce $\alpha$ for each $N_z$. The values of
$\alpha$ are shown in Table \ref{tab:criexp} for several $N_z$.

\begin{figure}
\centerline{\epsfig{file=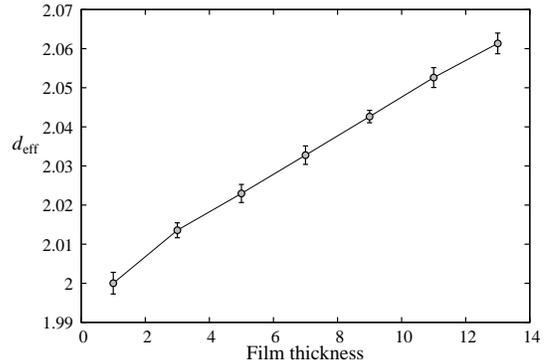,width=2.8in}}
\caption{Effective dimension of thin film obtained by using
effective exponents, as a function of thickness. } \label{de}
\end{figure}

It is interesting to note that the effective exponents obtained
above give rise to the effective dimension of thin film. This is
conceptually not rigorous but this is what observed in
experiments. Replacing the effective values of $\alpha$ obtained
above in $d_{\mbox{eff}}=(2-\alpha)/\nu$ we obtain
$d_{\mbox{eff}}$ shown in Fig. \ref{de}.

\begin{figure}
\centerline{\epsfig{file=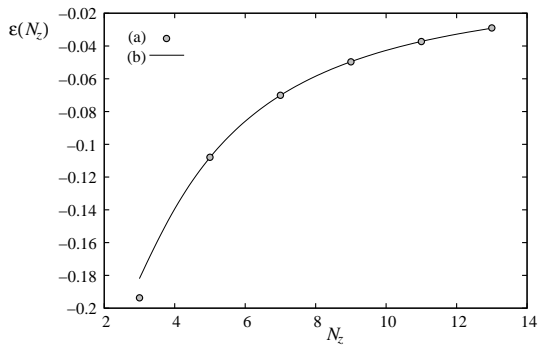,width=2.8in}} \caption{Critical
temperature at infinite $L$ as a function of the film thickness.
Points are MC results, continuous line is the prediction of
Capehart and Fisher, Eq. (\ref{CF}). The agreement is excellent. }
\label{TCINF}
\end{figure}

We note that $d_{\mbox{eff}}$ is very close to 2. It varies from 2
to $\simeq 2.061$ for $N_z$ going from 1 to 13. The 2D character
is thus dominant even with larger $N_z$. This supports the idea
that the finite correlation in the $z$ direction, though
qualitatively causing a deviation, cannot strongly alter  the 2D
critical behavior.  This point is interesting because, as said
earlier, some thermodynamic properties may show already their 3D
values at a thickness of about a few dozens of layers, but not the
critical behavior.  To show an example of this, let us plot in
Fig. \ref{TCINF} the transition temperature at $L=\infty$ for
several $N_z$, using Eq. \ref{TC} for each given $N_z$.  As seen,
$T_c(\infty)$ reaches already $\simeq 4.379$ at $N_z=13$ while its
value at 3D is $4.51$.\cite{Ferrenberg3,Blote}  A rough
extrapolation shows that the 3D values is attained for $N_z\simeq
25$ while the critical exponents at this thickness are far away
from  the 3D ones.

Let us show the prediction of Capehart and Fisher\cite{Fisher} on
the critical temperature as a function of $N_z$.
 Defining the critical-point shift as
\begin{equation}
\varepsilon(N_z)=\left[ T_c(L=\infty,N_z)-T_c(3D)\right]/T_c(3D)
\end{equation}
they showed that

\begin{equation}\label{CF}
\varepsilon(N_z)\approx \frac{b}{N_z^{1/\nu}}[1+a/N_z]
\end{equation}
where $\nu=0.6289$ (3D value). Using  $T_c(3D)=4.51$,  we fit the
above formula with $T_c(L=\infty,N_z)$ taken from Table
\ref{tab:criexp}, we obtain $a=-1.37572$ and $b=-1.92629$.  The MC
results and the fitted curve are shown in Fig. \ref{TCINF}. Note
that if we do not use the correction factor $[1+a/N_z] $, the fit
is not good for small $N_z$.  The prediction of Capehart and
Fisher is thus very well verified.

 We give here the precise values of $T_c(L=\infty, N_z)$ for each
thickness. For $N_z=1$, we have $T_c(L=\infty,N_z=1) = 2.2699\pm
0.0005$. Note that the exact value of $T_c(\infty)$ is
 2.26919 by solving the equation $\sinh^2(2J/T_c)=1$.  Again here, the
excellent
 agreement of our result shows the efficiency of the multiple histogram
technique as applied
 in the present paper.
The values of $T_c(L=\infty)$ for other $N_z$ are summarized in
Table~\ref{tab:criexp}.

\begin{figure}
\centerline{\epsfig{file=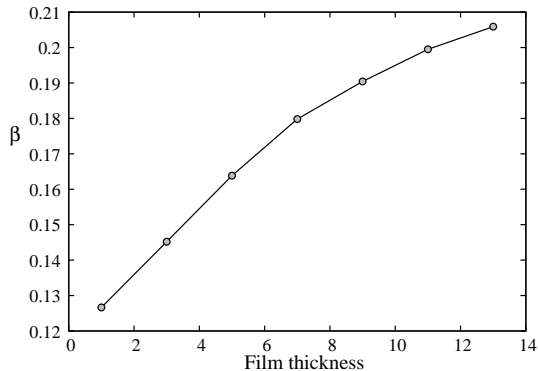,width=2.8in}}
\caption{Effective exponent $\beta$, obtained by using Eq.
\ref{MB}, versus the film thickness. } \label{BETA}
\end{figure}

Calculating now $M(L)$ at these values of $T_c(L=\infty,N_z)$ and
using Eq. \ref{MB}, we  obtain $\beta/\nu$ for each $N_z$.  For
$N_z = 1$, we have $\beta/\nu = 0.1268 \pm 0.0022$ which yields
$\beta = 0.1266\pm 0.0049$ which is in excellent agreement with
the exact result 0.125. Note that if we calculate $\beta$ from
$\alpha +2\beta +\gamma =2$, then $\beta = (2-1.75198-0.00199)/2 =
0.12302\pm 0.0035$ which is in good agreement with the direct
calculation within errors.  We show in Fig. \ref{BETA} the values
of $\beta$ obtained by direct calculation using Eq.~\ref{MB}. Note
that the deviation of $\beta$ from the 2D value when $N_z$ varies
from 1 to 13 is due to the crossover effect discussed in
subsection \ref{film}. It represents about 60$\%$. Remember that
the 3D value of $\beta$ is $0.3258\pm 0.0044$.\cite{Ferrenberg3}

Finally, for convenience, let us summarize our results in Table
\ref{tab:criexp} for $N_z=1,3,...,13$.  Except for $N_z=1$, all
other cases are effective exponents discussed above.  Due to the
smallness of $\alpha$, its value is shown with 5 decimals without
rounding.

\begin{widetext}
\begin{table*}
  \centering
  \caption{Critical exponents, effective dimension and critical temperature
  at infinite $xy$ limit as obtained in this paper.}\label{tab:criexp}
  \begin{tabular}{| r | c | c | c | c | c | c |}
    \hline
    $N_z$ & $\nu$ & $\gamma$ & $\alpha$ & $\beta$ & $d_{\mathrm{eff}}$ &
$T_c(L=\infty,N_z)$ \\
\hline 1 & $0.9990 \pm 0.0028$ & $1.7520 \pm 0.0062$ & $0.00199
\pm 0.00279$ & $0.1266 \pm 0.0049$ & $2.0000 \pm 0.0028$ &
$2.2699\pm
0.0005$ \\
3 & $0.9922 \pm 0.0019$ & $1.7377 \pm 0.0035$ & $0.00222 \pm
0.00192$ & $0.1452 \pm 0.0040$ & $2.0135 \pm 0.0019$ & $3.6365 \pm
0.0024$ \\
5 & $0.9876 \pm 0.0023$ & $1.7230 \pm 0.0069$ & $0.00222 \pm
0.00234$ & $0.1639 \pm 0.0051$ & $2.0230 \pm 0.0023$ & $4.0234 \pm
0.0028$ \\
7 & $0.9828 \pm 0.0024$ & $1.7042 \pm 0.0087$ & $0.00223 \pm
0.00238$ & $0.1798 \pm 0.0069$ & $2.0328 \pm 0.0024$ & $4.1939 \pm
0.0032$ \\
9 & $0.9780 \pm 0.0016$ & $1.6736 \pm 0.0084$ & $0.00224 \pm
0.00161$ & $0.1904 \pm 0.0071$ & $2.0426 \pm 0.0016$ & $4.2859 \pm
0.0022$ \\
11& $0.9733 \pm 0.0025$ & $1.6354 \pm 0.0083$ & $0.00224 \pm
0.00256$ & $0.1995 \pm 0.0088$ & $2.0526 \pm 0.0026$ & $4.3418 \pm
0.0032$ \\
13& $0.9692 \pm 0.0026$ & $1.6122 \pm 0.0102$ & $0.00226 \pm
0.00268$ & $0.2059 \pm 0.0092$ & $2.0613 \pm 0.0027$ & $4.3792 \pm
0.0034$ \\
    \hline
  \end{tabular}
\end{table*}
\end{widetext}

\subsection{Larger sizes and correction to scaling}

We consider here the effects of larger $L$ and of the correction
to scaling. For the effect of larger $L$, we will  extend our size
up to $L=160$, for just the case  $N_z=3$.

\begin{figure}
\centerline{\epsfig{file=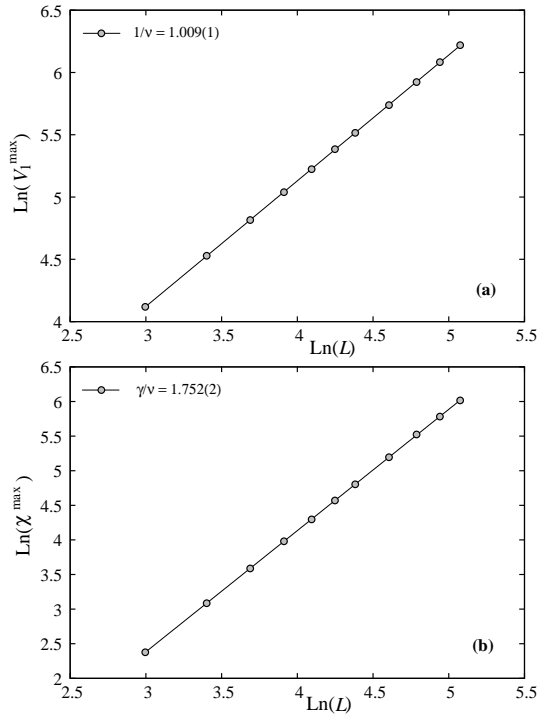,width=2.8in}} \caption{(a)
$V_1^{max}$ and (b) $\chi^{max}$ vs $L$ up to 160 with $N_z=3$.}
\label{fig:GNL160}
\end{figure}

The results indicate that larger $L$ does not change the results
shown above. Figure \ref{fig:GNL160}(a) displays the maximum of
$V_1$ as a function of $L$ up to 160. Using Eq. (\ref{V1}), i. e.
without correction to scaling, we obtain $1/\nu=1.009\pm0.001$
which is to be compared to $1/\nu=1.008\pm0.002$ using $L$ up to
80. The change is therefore insignificant because it is at the
third decimal i. e. at the error level.  The same is observed for
$\gamma/\nu$ as shown in Fig. \ref{fig:GNL160}(b): $ \gamma/\nu =
1.752\pm 0.002$ using $L$ up to 160 instead of $ \gamma/\nu =
1.751\pm 0.002$ using $L$ up to 80.

\begin{figure}
\centerline{\epsfig{file=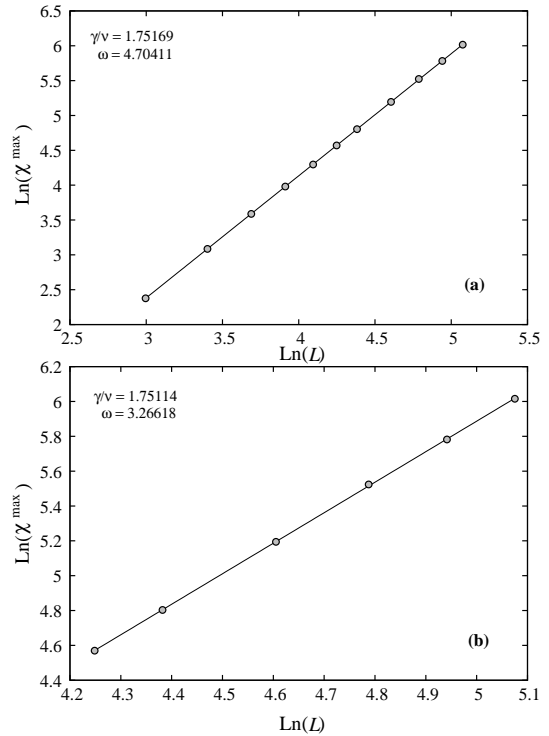,width=2.8in}}
\caption{$\chi^{max}$ vs $L$ (a) from $20$ up to 160  (b) from
$70$ up to 160, for $N_z=3$.} \label{fig:CORRECT}
\end{figure}

Now, let us allow for correction to scaling, i. e. we use
Eq.(\ref{chic}) instead of Eq. (\ref{chis}) for fitting. We obtain
the following values:  $\gamma/\nu=1.751\pm0.002$, $B_1= 0.05676$,
$B_2=1.57554 $, $\omega=3.26618$ if we use $L$ = 70 to 160 (see Fig.
\ref{fig:CORRECT}). The value of $\gamma/\nu$ in the case of no
scaling correction is $1.752\pm0.002$. Therefore, we can conclude
that this correction is insignificant. The large value of $\omega$
explains the smallness of the correction.

\subsection{Role of boundary condition}

\begin{figure}
\centerline{\epsfig{file=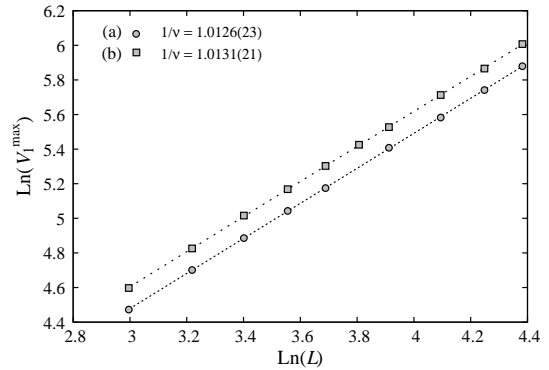,width=2.8in}} \caption{Maximum
of the first derivative of $\ln M$  versus $ L$ in the $\ln-\ln$
scale for $N_z=5$ (a) without PBC in $z$ direction (b) with PBC in
$z$ direction. The slopes are indicated on  the figure.  See text
for comments. } \label{fig:NUZ5}
\end{figure}

\begin{figure}
\centerline{\epsfig{file=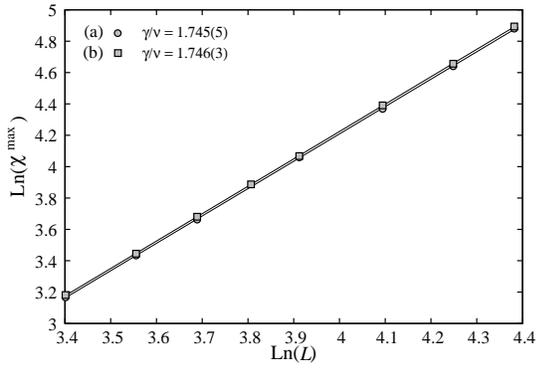,width=2.8in}}
\caption{Maximum of susceptibility versus $L$ in the $\ln-\ln$
scale for $N_z=5$ (a) without PBC in $z$ direction (b) with PBC in
$z$ direction. The points of these cases cannot be distinguished
in the figure scale.  The slopes are indicated on  the figure.
See text for comments.} \label{fig:GAMZ5}
\end{figure}

\begin{figure}
\centerline{\epsfig{file=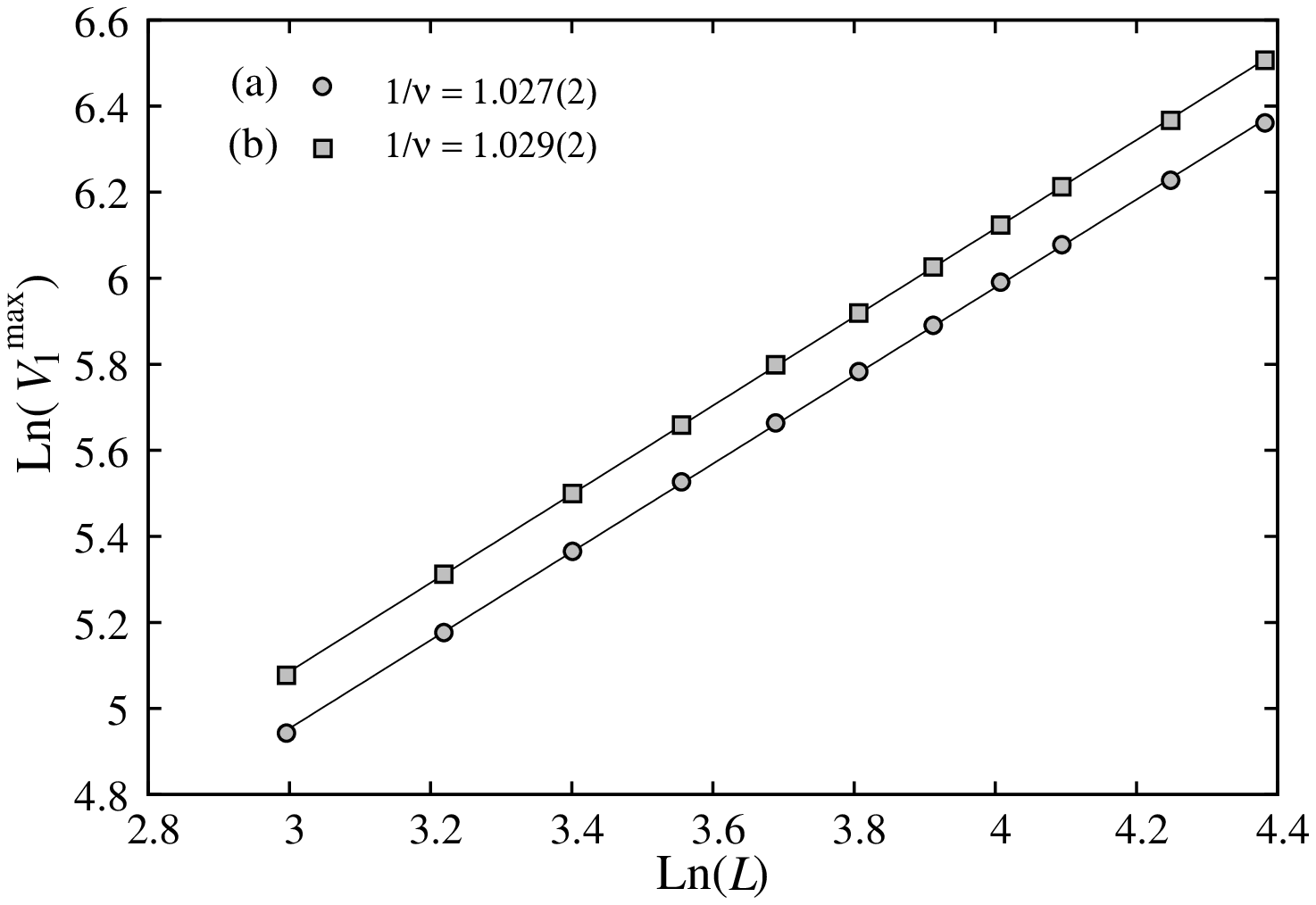,width=2.8in}}
\caption{Maximum of the first derivative of $\ln M$  versus $ L$
in the $\ln-\ln$ scale for $N_z=11$ (a) without PBC in $z$
direction (b) with PBC in $z$ direction.  The slopes are indicated
on  the figure.  See text for comments. } \label{fig:NUZ11}
\end{figure}

\begin{figure}
\centerline{\epsfig{file=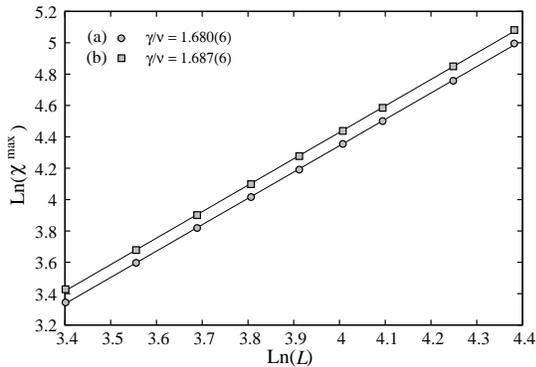,width=2.8in}}
\caption{Maximum of susceptibility versus $L$ in the $\ln-\ln$
scale for $N_z=11$ (a) without PBC in $z$ direction (b) with PBC
in $z$ direction. The slopes are indicated on  the figure.  See
text for comments. } \label{fig:GAMZ11}
\end{figure}

\begin{figure}
\centerline{\epsfig{file=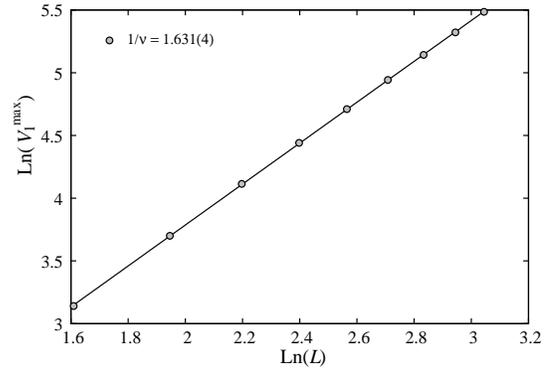,width=2.8in}} \caption{Maximum
of the first derivative of $\ln M$  versus $ L$ in the $\ln-\ln$
scale for 3D case.} \label{fig:NUL3D}
\end{figure}

\begin{figure}
\centerline{\epsfig{file=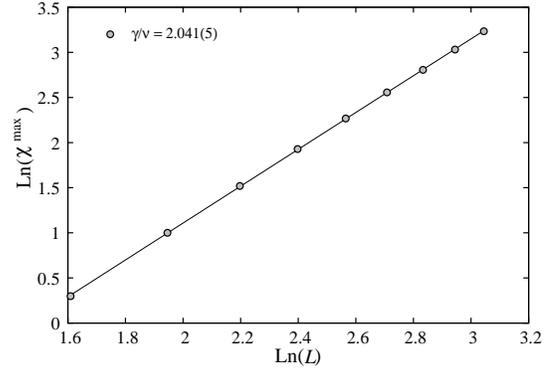,width=2.8in}}
\caption{Maximum of susceptibility versus $L$ in the $\ln-\ln$
scale for 3D case.} \label{fig:GAML3D}
\end{figure}

To close this section, let us touch upon the question: does the
absence of PBC in the $z$ direction cause the deviation of the
critical exponents?  The answer is no: we have calculated $\nu$ and
$\gamma$ for $N_z=5$ in both cases, with and without PBC in the $z$
direction. The results show no significant difference between the
two cases as seen in Figs. \ref{fig:NUZ5} and \ref{fig:GAMZ5}. We
have found the same thing with $N_z=11$ shown in Figs.
\ref{fig:NUZ11} and \ref{fig:GAMZ11}. So, we conclude that the fixed
thickness will result in the deviation of the critical exponents,
not from the absence of the PBC. This is somewhat surprising since
we may think, incorrectly, that the PBC should mimic the infinite
dimension so that we should obtain the 3D behavior when applying the
PBC.  As will be seen below, the 3D behavior is recovered only when
the finite size scaling is applied in the $z$ direction at the same
time in the $xy$ plane.  To show this, we plot in Figs.
\ref{fig:NUL3D} and \ref{fig:GAML3D} the results for the 3D case.
Even with our modest sizes (up to $L=N_z=21$, since it is not our
purpose to treat the  3D case here), we obtain $\nu=0.613\pm0.005$
and $\gamma=1.250\pm 0.005$ very close to their 3D best known values
$\nu_{3D}=0.6302\pm0.0001$ from Ref. \onlinecite{Blote} and
$\nu_{3D}=0.6289\pm0.0008$ and $\gamma_{3D}=1.2390\pm 0.0025$
obtained by using  $24\leq L \leq 96$ given in Ref.
\onlinecite{Ferrenberg3}.

\section{Concluding remarks}

We have considered a simple system, namely the Ising model on a
simple cubic thin film,  in order to clarify the point whether or
not there is a continuous deviation of the 2D exponents with varying
film thickness.  From results obtained by the highly accurate
multiple histogram technique shown above, we conclude that the
critical exponents in thin films show a continuous deviation from
their 2D values as soon as the thickness departs from 1. This
deviation stems from a deep physical mechanism: Capehart and
Fisher\cite{Fisher} have argued that if one works exactly at the
critical temperature $T_c(L=\infty,N_z)$ then the critical exponents
should be those of 2D universality class as long as the film
thickness is finite. At $T_c(L=\infty,N_z)$, the correlation in the
$z$ direction $\xi$ remains finite while those in the $xy$ planes
become infinite.  Hence $\xi$ is irrelevant to the criticality. This
yields therefore the 2D behavior.  However, when the system is away
from $T_c(L=\infty,N_z)$,  as is the case in numerical simulations
using finite sizes, the system may have a 3D behavior as long as
$\xi<<N_z$.  This should yield a deviation of 2D critical exponents.
The results we obtained in this paper verify this picture.  In
addition, the prediction of Capehart and Fisher for the shift of the
critical temperature with the film thickness is in a perfect
agreement with our simulations. Note furthermore  that (i) the
deviations of the exponents from their 2D values are very different
in magnitude: while $\nu$ and $\alpha$ vary very little over the
studied range of thickness, $\gamma$ and specially $\beta$ suffer
stronger deviations, (ii) with a fixed thickness $N_z \neq 1$,   the
same "effective" exponents are observed, within errors, in
simulations with and without periodic boundary condition in the $z$
direction, (iii) to obtain the 3D behavior, the finite size scaling
should be applied simultaneously in the three directions, i. e. all
dimensions should be allowed to go to infinity. If we do not apply
the scaling in the $z$ direction, we will not obtain 3D behavior
even with a very large, but fixed, thickness and even with periodic
boundary condition in the $z$ direction, (iv) with regard to the
critical behavior, thin films behave as systems with effective
critical exponents whose values are those between 2D and 3D.

To conclude, we hope that the numerical results shown in this
paper will  help experimentalists to interpret their data which
are usually obtained at a finite distance from the critical point.
It should be also desirable to study more cases to clarify the
role of thickness on the  behavior of very thin films, in
particular the effect of the film thickness on the bulk
first-order transition.

One of us (VTN) thanks the "Asia Pacific Center for Theoretical
Physics" (South Korea) for a financial post-doc support and
hospitality during the period 2006-2007 where part of this work
was carried out.  The authors are grateful to Yann Costes of the
University of Cergy-Pontoise for technical help in parallel
computation.

{}

\end{document}